\documentclass[12pt,a4]{article}
\usepackage{sw20lart}

\input tcilatex
\QQQ{Language}{
American English
}

\begin{document}

\author{C.A.G. Sasaki, \\
UERJ, Universidade Estadual do Rio de Janeiro;\\
Departamento de Estruturas Matem\'{a}ticas; \\
IME, Instituto de Matem\'{a}tica; \\
Rua S\~{a}o Francisco Xavier 524, 20550-013, \\
Maracan\~{a}, Rio de Janeiro, Brazil. \thinspace \and D.G.G. Sasaki \\
CBPF, Centro Brasileiro de Pesquisas F\'{\i}sicas,\\
DCP, Departamento de Campos e Part\'{\i}culas.\\
Rua Dr Xavier Sigaud 150, 22290-180 Urca,\\
Rio de Janeiro, Brazil.\\
and \and S.P. Sorella \\
UERJ, Universidade Estadual do Rio de Janeiro,\\
Departamento de F\'{\i}sica Te\'{o}rica; \\
IF, Instituto de F\'{\i}sica; \\
Rua S\~{a}o Francisco Xavier 524, 20550-013; \\
Maracan\~{a}, Rio de Janeiro, Brazil. \and \textbf{UERJ/DFT-04/98}%
\vspace{2mm}\newline \and \textbf{PACS: 11.10.Gh}}
\title{\textbf{Nonlinear vector susy for the three-dimensional topological massive
Yang-Mills theory }}
\maketitle

\begin{abstract}
A nonlinear vector supersymmetry for three-dimensional topological massive
Yang-Mills is obtained by making use of a nonlinear but local and covariant
redefinition of the gauge field.
\end{abstract}

\section{Introduction}

Nowadays, it is well established that the topological field theories (TFT) 
\cite{alg2} are characterized, besides their BRST invariance, by a further
symmetry carrying a Lorentz index \cite{alg3,alg4,alg5,alg6}. This
additional invariance has been called vector supersymmetry since the
corresponding generators give rise, together with the BRST generator, to a
supersymmetric algebra of the Wess-Zumino type. The existence of the vector
supersymmetry has been first detected in the case of the three-dimensional
topological Chern-Simons theory quantized in the Landau gauge \cite
{alg3,alg4} and later on has been extended to others TFT as, for instance,
the BF models \cite{alg5,book} and the Witten's cohomological field theories 
\cite{alg6}. It is worth remarking here that, actually, the existence of the
vector susy can be established by purely cohomological methods, relying on
the fact that the energy-momentum tensor of the TFT can be set in the form
of an exact BRST variation \cite{algebraic}.

On the other side, in a recent series of works \cite{simple,chern}, it has
been proven that three-dimensional gauge theories of the Yang-Mills type in
presence of the topological Chern-Simons term can be cast in the form of a
pure Chern-Simons action through a nonlinear but local and covariant
redefinition of gauge field. Indeed, in the case of the topological massive
Yang-Mills theory whose action is given by the sum of the Chern-Simons
action and of the Yang-Mills term

\begin{equation}
S_{TMYM}\left( A\right) =S_{CS}\left( A\right) +S_{YM}\left( A\right) ,
\label{1.1}
\end{equation}
where 
\begin{equation}
S_{YM}\left( A\right) =\frac 1{4m}tr\int d^3x\,\,F_{\mu \nu }F^{\mu \nu },
\label{YMACT}
\end{equation}
\begin{equation}
S_{CS}\left( A\right) =\frac 12tr\int d^3x\,\,\varepsilon ^{\mu \nu \rho
}\left( A_\mu \partial _vA_\rho +\frac 23gA_\mu A_\nu A_\rho \right) ,
\label{CSACT}
\end{equation}
we have

\begin{equation}
S_{TMYM}\left( A\right) =S_{CS}(\widehat{A})\;,  \label{1.2}
\end{equation}
with

\begin{equation}
\widehat{A}_\mu =A_\mu +\sum_{n=1}^\infty \,\frac 1{m^n}\vartheta _\mu
^n\left( D,F\right) \;.  \label{1.3}
\end{equation}
The parameters $g$,$m$ in eqs.$\left( \text{\ref{YMACT}}\right) ,\left( 
\text{\ref{CSACT}}\right) $ identify respectively the coupling constant and
the so called topological mass \cite{djt}. As shown in ref.\cite{simple},
the coefficients $\vartheta _\mu ^n$ turn out to be local and covariant,
meaning that they can be expressed in terms of the field strength $F_{\mu v}$
and its covariant derivatives. For instance, the first terms of the
redefinition $\left( \text{\ref{1.3}}\right) $ have been found 
\begin{eqnarray}
\vartheta _\mu ^1 &=&\frac 14\varepsilon _{\mu \sigma \tau }F^{\sigma \tau },
\nonumber  \label{1.4} \\
\vartheta _\mu ^2 &=&\frac 18D^\sigma F_{\sigma \mu }\;,  \nonumber \\
\vartheta _\mu ^3 &=&-\frac 1{16}\varepsilon _{\mu \sigma \tau }D^\sigma
D_\rho F^{\rho \tau }+\frac g{48}\varepsilon _{\mu \sigma \tau }\left[
F^{\sigma \rho },F_\rho ^{\,\,\,\tau }\right] \;,  \nonumber \\
\vartheta _\mu ^4 &=&-\frac 5{128}D^2D^\rho F_{\rho \mu }+\frac 5{128}D^\nu
D_\mu D^\lambda F_{\lambda \nu }  \nonumber \\
&&-\frac 7{192}g\left[ D^\rho F_{\rho \tau },F_\mu ^{\,\,\,\tau }\right]
-\frac g{48}\left[ D_\nu F_{\mu \lambda },F^{\lambda \nu }\right] \;.
\label{1.4}
\end{eqnarray}
The equation $\left( \text{\ref{1.2}}\right) $ expresses the classical
equivalence, up to a field redefinition, between the topological massive
Yang-Mills and the pure Chern-Simons action.

It seems therefore natural to ask ourselves if, due to the equation $\left( 
\text{\ref{1.2}}\right) $, the vector supersymmetry may still be present
when the Yang-Mills action is added to the pure topological Chern-Simons
term. This is the purpose of the present letter. We shall be able, in
particular, to prove that the topological massive Yang-Mills action $\left( 
\text{\ref{1.1}}\right) $ does posses in fact the vector supersymmetry.
Furthermore, unlike the pure Chern-Simons case, the vector supersymmetry is
now realized nonlinearly on the fields and, in analogy with the field
redefinition $\left( \text{\ref{1.3}}\right) $, can be cast in the form of a
power series in $1/m$.

The work is organized as follows. In Sect.2 we briefly recall the
supersymmetric structure of pure Chern-Simons. The Sect.3 is devoted to the
analysis of the vector susy in the case of the topological massive
Yang-Mills theory.

\section{Vector susy in the pure Chern-Simons theory}

In order to recall the main features of the vector susy in pure Chern-Simons
theory \cite{book,algebraic}, we first quantize the model by adopting a
transverse Landau gauge condition. For the gauge fixed action we get

\begin{equation}
\Sigma _{CS}\left( A\right) =S_{CS}\left( A\right) +tr\int d^3x\left(
b\partial ^\mu A_\mu +\overline{c}\partial ^\mu D_\mu c\right) ,  \label{2.1}
\end{equation}
where the fields $b,c,\overline{c}$ denote respectively that Lagrange
multiplier and the Faddeev-Popov ghosts. For the BRST transformations we have

\begin{eqnarray}
sA_\mu &=&-D_\mu c=-\left( \partial _\mu c+g\left[ A_\mu ,c\right] \right) ,
\nonumber  \label{2.4} \\
sc &=&gc^2,  \nonumber \\
s\overline{c} &=&b\,,  \nonumber  \label{2.2} \\
sb &=&0\,,  \label{2.2}
\end{eqnarray}
with

\begin{equation}
s\Sigma _{CS}=0\,.  \label{2.3}
\end{equation}
It is easily checked now that the quantized action $\Sigma _{CS\text{ }}$is
left invariant by the following vector type transformations:

\begin{eqnarray}
\delta _\mu c &=&A_\mu \,,  \nonumber \\
\delta _\mu A_v &=&\varepsilon _{\mu v\rho }\partial ^\rho \overline{c\,}, 
\nonumber \\
\delta _\mu b &=&\partial _\mu \overline{c\,},  \nonumber \\
\delta _\mu \overline{c} &=&0\,,  \label{2.4}
\end{eqnarray}
with

\begin{equation}
\delta _\mu \Sigma _{CS}=tr\int d^3x\left( A_\mu \frac \delta {\delta
c}+\varepsilon _{\mu v\rho }\partial ^\rho \overline{c}\frac \delta {\delta
A_v}+\partial _\mu \overline{c}\frac \delta {\delta b}\right) \Sigma
_{CS}=0\,.  \label{2.5}
\end{equation}
In addition, the generators $\delta _\mu ,s$ give rise to the following
anti-commutation relations

\begin{eqnarray}
\left\{ \delta _\mu ,\delta _v\right\} &=&0\;,  \nonumber \\
\left\{ s,\delta _\mu \right\} &=&\partial _\mu +\left( \text{equations of
motion}\right) \;,  \label{2.6}
\end{eqnarray}
which, closing on-shell on the translations, yield a Wess-Zumino type
supersymmetric algebra. The eqs.$\left( \ref{2.4}\right) $ are known as the
vector susy transformations, since the generator $\delta _\mu $ carries a
Lorentz index. It is worth noticing here that the vector susy $\left( \ref
{2.4}\right) $ is linearly realized on the fields and that it has been
proven to play an important role on the proof of the ultraviolet finiteness
of the Chern-Simons theory \cite{alg4,book}.

\section{The case of topological massive Yang-Mills}

In order to discuss the existence of the vector susy for the topological
massive Yang-Mills action $\left( \ref{1.1}\right) $, we remind that the
coefficients $\vartheta _\mu ^n$ in the equations $\left( \ref{1.3}\right) $%
, $\left( \ref{1.4}\right) $ have been proven \cite{chern} to transform
covariantly under the BRST transformations, \textit{i.e.}

\begin{equation}
s\vartheta _\mu ^n=g\left[ \vartheta _\mu ^n,c\right] \;.  \label{3.1}
\end{equation}
As a consequence, the redefined field $\widehat{A}_\mu $ transforms as a
connection, namely

\begin{equation}
s\widehat{A}_\mu =-\left( \partial _\mu c+g\left[ \widehat{A}_\mu ,c\right]
\right) \;.  \label{3.2}
\end{equation}
Therefore, if we choose as the gauge fixing condition a Landau condition for
the connection $\widehat{A}_\mu $, \textit{i. e.}

\begin{equation}
\partial ^\mu \widehat{A}_\mu =0\;,  \label{3.3}
\end{equation}
for the quantized topological massive Yang-Mills action we get:

\begin{eqnarray}
\Sigma _{TMYM}\left( A\right) &=&S_{TMYM}\left( A\right)  \nonumber \\
&&+tr\int d^3x\left( b\partial ^\mu \widehat{A}_\mu +\overline{c}\partial
^\mu \left( \partial _\mu c+g\left[ \widehat{A}_\mu ,c\right] \right)
\right) \;.  \label{3.4}
\end{eqnarray}
Moreover, recalling from eq.$\left( \text{\ref{1.2}}\right) $ that

\begin{equation}
S_{TMYM}\left( A\right) =S_{CS}\left( A\right) +\frac 1{4m}tr\int
d^3xF^2=S_{CS}\left( \widehat{A}\right) ,  \label{3.5}
\end{equation}
the equation $\left( \text{\ref{3.4}}\right) $ becomes

\begin{eqnarray}
\Sigma _{TMYM}\left( A\right) &=&\Sigma _{CS}(\widehat{A})  \label{3.6} \\
&=&S_{TMYM}(\widehat{A})+tr\int d^3x\left\{ b\partial ^\mu \widehat{A}_\mu +%
\overline{c}\partial ^\mu \left( \partial _\mu c+g\left[ \widehat{A}_\mu
,c\right] \right) \right\} \;.  \nonumber
\end{eqnarray}
The expression $\left( \text{\ref{3.6}}\right) $ is easily recognized to be
the quantized Chern-Simons action $\left( \text{\ref{2.1}}\right) $ viewed
as a functional of the gauge connection $\widehat{A}_\mu $ instead of $A_\mu 
$. It is apparent therefore that the quantized action $\Sigma _{TMYM}\left(
A\right) $ is left invariant by the following vector transformations

\begin{equation}
tr\int d^3x\left( \widehat{A}_\mu \frac \delta {\delta c}+\varepsilon _{\mu
v\rho }\partial ^\rho \overline{c}\frac \delta {\delta \widehat{A}%
_v}+\partial _\mu \overline{c}\frac \delta {\delta b}\right) \Sigma
_{TMYM}\left( A\right) =0\;.  \label{3.7}
\end{equation}
Moving now from the connection $\widehat{A}_\mu $ to the gauge field $A_\mu $%
, we obtain the nonlinear vector susy Ward identity for the quantized
topological massive Yang-Mills action we were looking for 
\begin{equation}
\mathcal{W}_\mu \Sigma _{TMYM}=0\;,  \label{3.8}
\end{equation}
\[
\mathcal{W}_\mu =tr\int d^3x\;\left( (A_\mu +\sum_{n=1}^\infty \,\frac
1{m^n}\vartheta _\mu ^n)\frac \delta {\delta c}+\varepsilon _{\mu \alpha
\beta }\partial ^\alpha \overline{c}\mathcal{M}_\lambda ^\beta (x)\frac
\delta {\delta A_\lambda }+\partial _\mu \overline{c}\frac \delta {\delta
b}\right) \;,
\]
where the kernel $\mathcal{M}_\lambda ^\beta (x)$ is given by

\begin{equation}
\mathcal{M}_\lambda ^\beta (x)=\int d^3y\frac{\delta A_\lambda (y)}{\delta 
\widehat{A}_\beta \left( x\right) }\;,  \label{inv}
\end{equation}
and is easily obtained by inverting the transformation $\left( \text{\ref
{1.3}}\right) .$ We see thus that, as already mentioned, the vector susy
Ward identity for the massive topological Yang-Mills theory is realized
nonlinearly, due to the presence of the coefficients $\vartheta _\mu ^n$ and
of the kernel $\mathcal{M}_\lambda ^\beta (x)$ in the Ward operator $%
\mathcal{W}_\mu $. Furthermore, as in the case of the field redefinition $%
\left( \text{\ref{1.3}}\right) $, the vector susy Ward operator $\mathcal{W}%
_\mu $ can be expanded in a power series in $1/m$, yielding

\begin{equation}
\mathcal{W}_\mu =\sum_{n=0}^\infty \frac 1{m^n}\mathcal{W}_\mu ^n\;.
\label{3.9}
\end{equation}
For the first terms of the series we get 
\begin{eqnarray}
\mathcal{W}_\mu ^0 &=&tr\int \left( A_\mu \frac \delta {\delta
c}+\varepsilon _{\mu \alpha \beta }\partial ^\beta \overline{c}\frac \delta
{\delta A_\alpha }+\partial _\mu \overline{c}\frac \delta {\delta b}\right)
\;,  \label{w-tr} \\
\mathcal{W}_\mu ^1 &=&tr\int \left( \vartheta _\mu ^1\frac \delta {\delta
c}+\frac 12D_\rho \partial ^\rho \overline{c}\frac \delta {\delta A^\mu
}-\frac 12D_\mu \partial ^\rho \overline{c}\frac \delta {\delta A^\rho
}\right) \;,  \nonumber \\
\mathcal{W}_\mu ^2 &=&tr\int \left( \vartheta _\mu ^2\frac \delta {\delta
c}+\frac 18\varepsilon _{\mu \nu \rho }\left( 3D^2\partial ^\rho \overline{c}%
\delta ^{\nu \alpha }-3D^\nu D^\alpha \partial ^\rho \overline{c}-\left[
\partial ^\rho \overline{c},F^{\nu \alpha }\right] \right) \frac \delta
{\delta A^\alpha }\right) \;,  \nonumber \\
\mathcal{W}_\mu ^3 &=&tr\int (\vartheta _\mu ^3\frac \delta {\delta c}+\frac
12\left( \frac 14D_\nu \left[ \partial ^\rho \overline{c},F_{\,\,\,\rho
}^\nu \right] +\frac 58D^2D_\rho \partial ^\rho \overline{c}-\left[ \partial
^\rho \overline{c},\vartheta _\rho ^2\right] \right) \frac \delta {\delta
A^\mu }  \nonumber \\
&&+\frac 12(\frac 38\left[ \partial ^\nu \overline{c},D^\rho F_{\nu \mu
}\right] -\left[ \partial ^\rho \overline{c},\vartheta _\mu ^2\right] -\frac
98D_\mu \left[ \partial ^\nu \overline{c},F_{\,\,\nu }^\rho \right] 
\nonumber \\
&&-\frac 98D_\nu \left[ \partial ^\nu \overline{c},F_\mu ^{\,\,\,\rho
}\right] -\frac 54D^\rho \left[ \partial ^\nu \overline{c},F_{\nu \mu
}\right] +\frac 14D^\nu \left[ \partial ^\rho \overline{c},F_{\mu \nu
}\right]  \nonumber \\
&&+\frac 58\left( D_\nu D^\rho D_\mu \partial ^\nu \overline{c}-D^2D_\mu
\partial ^\rho \overline{c}-D_\mu D^\rho D_\nu \partial ^\nu \overline{c}%
\right) )\frac \delta {\delta A^\rho })\;.  \nonumber
\end{eqnarray}
Notice also that $\mathcal{W}_\mu ^0$ in the eqs.$\left( \text{\ref{w-tr}}%
\right) $ coincides with the vector susy Ward operator $\delta _\mu $ $%
\left( \ref{2.5}\right) $ for the pure Chern-Simons. Finally, according to
eqs. $\left( \text{\ref{2.6}}\right) $, for the algebra generated by the
BRST operator and by the vector susy Ward operator $\mathcal{W}_\mu $ we
obtain the following on-shell Wess-Zumino type algebra

\begin{eqnarray}
\left\{ s,\mathcal{W}_\mu \right\} A_v &=&\partial _\mu A_v+2\varepsilon
_{v\mu \rho }\int d^3y\frac{\delta A_\lambda (y)}{\delta \widehat{A}_\rho
\left( x\right) }\frac{\delta \Sigma _{TMYM}\left( A\right) }{\delta
A_{\lambda (}y)}\;,  \nonumber \\
\left\{ s,\mathcal{W}_\mu \right\} c &=&\partial _\mu c\;,  \nonumber \\
\left\{ s,\mathcal{W}_\mu \right\} \overline{c} &=&\partial _\mu \overline{c}%
\;,  \nonumber \\
\left\{ s,\mathcal{W}_\mu \right\} b &=&\partial _\mu b\;.  \label{3.11}
\end{eqnarray}
\newpage\ 

{\Large \textbf{Acknowledgements}}

The Conselho\ Nacional de Pesquisa e Desenvolvimento (CNPq /Brazil), the
Faperj, Funda\c {c}\~{a}o de Amparo \`{a} Pesquisa do Estado do Rio de
Janeiro and the SR2-UERJ are gratefully acknowledged for financial support.

\end{document}